\documentstyle[aps,twocolumn,epsf]{revtex}

\def\IN{\relax{\rm I\kern-.18em N}}
\def\IR{\relax{\rm I\kern-.18em R}}

\font\cmss=cmss12 \font\cmsss=cmss12 at 7pt
\def\IZ{\relax\ifmmode\mathchoice
{\hbox{\cmss Z\kern-.4em Z}}{\hbox{\cmss Z\kern-.4em Z}}
{\lower.9pt\hbox{\cmsss Z\kern-.4em Z}}
{\lower1.2pt\hbox{\cmsss Z\kern-.4em Z}}\else{\cmss Z\kern-.4em Z}\fi}

\def\inbar{\,\vrule height1.5ex width.4pt depth0pt}
\def\IC{\relax\hbox{$\inbar\kern-.3em{\rm C}$}}

\newcommand{\Od}{{\cal O}}

\newcommand{\Tr}{\mbox{Tr}}

\newcommand{\pabar}{\not{\!{\partial}}}

\newcommand{\Qbar}{\not{\!Q}}

\newcommand{\intT}{\int_T d^2x}
 
\newcommand{\pib}{\int_{periodic} \!\!\!\!\!\!\!\!\!\!d\phi \ }
\newcommand{\pif}{\int_{antiper} \!\!\!\!\!\!\!\!\!\!d\bar\psi d\psi}

\newcommand{\intk}{\int_{-\infty}^{+\infty} \frac{dk}{2\pi}}
\newcommand{\curr}{j_\mu (x)}

\newcommand{\prodj}{\prod_{j=1}^n}
\newcommand{\prodjN}{\prod_{j=1}^N}
\newcommand{\prodjint}{\prodj\int_T d^2 x_j d^2 y_j}

\newcommand{\sumn}{\sum_{n=0}^{\infty}}
\newcommand{\sigpm}{\sigma_\pm (x)}

\newcommand{\be}{\begin{equation}}
\newcommand{\ee}{\end{equation}}
\newcommand{\ba}{\begin{eqnarray}}
\newcommand{\ea}{\end{eqnarray}}
\newcommand{\dle}[1]{\label{#1}}
\newcommand{\dla}[1]{\label{#1}}
\newcommand{\dr}[1]{\ref{#1}}
\newcommand{\dc}[1]{\cite{#1}}
\newcommand{\dnote}[1]{}

\newcommand{\dbibitem}[1]{\bibitem{#1}} 

\newcommand{\gsim}{\raise.3ex\hbox{$>$\kern-.75em\lower1ex\hbox{$\sim$}}}
\newcommand{\lsim}{\raise.3ex\hbox{$<$\kern-.75em\lower1ex\hbox{$\sim$}}}

\newcommand{\half}{{\frac{1}{2}}}
\newcommand{\paa}{\partial}

\newcommand{\NP}[1]{{\it Nucl.\ Phys.\ }{\bf #1}}

\newcommand{\PL}[1]{{\em Phys.\ Lett.\ }{\bf #1}}

\newcommand{\CMP}[1]{{\em Comm.\ Math.\ Phys.\ }{\bf #1}}

\newcommand{\PR}[1]{{\em Phys.\ Rev.\ }{\bf #1}}

\begin{document}

\typeout{--- Title page start ---}

\renewcommand{\thefootnote}{\fnsymbol{footnote}}


\title{ Kinks versus fermions, \\
or \\
the 2D sine-Gordon versus massive
Thirring models,\\at $T>0$ and $\mu \neq 0$\thanks{Proceedings
of the talk based on [1] and given by
D.A.Steer at the 5th International Workshop on Thermal Field Theory,
Regensburg, Germany, August 10-14, 1998.}}

\author{A.G\'{o}mez Nicola$^{a}$\thanks{Gomez@eucmax.sim.ucm.es}, 
R.J.Rivers$^{b}$\thanks{R.Rivers@ic.ac.uk} and
D.A.Steer$^{c}$\thanks{D.A.Steer@damtp.cam.ac.uk} }

\address{a) Departamento de F\'{\i}sica 
Te\'orica, Universidad Complutense, 28040, Madrid, Spain. }

\address{ b) Theoretical Physics, 
Blackett Laboratory, Imperial College, Prince Consort Road, London,
SW7 2BZ, U.K.}

\address{ c) D.A.M.T.P., Silver Street, Cambridge, CB3 9EW, U.K.}

\date{\today}

\maketitle

\renewcommand{\thefootnote}{\arabic{footnote}}
\setcounter{footnote}{0}
\typeout{--- Main Text Start ---}

\begin{abstract}

The two dimensional (1+1) sine-Gordon model finds many applications in
condensed matter physics.  These in
turn provide an experimental means for the study of topological defects,
some of which may have had a huge impact on the early universe.  As a
first step in trying to exploit this analogy and also others
which exist with low-energy QCD, we study bosonisation in the massive
Thirring and sine-Gordon models at finite temperature
$T$ and nonzero fermion chemical potential $\mu$.  
Both canonical operator and path integral approaches are used to prove
the equality of the partition functions of the two models at $T>0$ and
$\mu=0$, as was recently shown.  This enables the
relationship between thermal normal ordering and path-integral
renormalisation to be specified.  Furthermore, we prove that
thermal averages of zero-charge  operators can
be identified as long as one uses the usual $T=0$ identification
between coupling constants.  Analysis of the point-split regularised 
fermion current then leads to the thermal equivalence between
sine-Gordon kinks and Thirring fermions.  At  $\mu\neq 0$ and $T>0$, 
we show, in perturbation theory around the massless Thirring model, 
 that the bosonised theory is the
sine-Gordon model plus an additional topological term which accounts
for the existence of net fermion charge excitations (the fermions or
the kinks) in the thermal bath. This result generalises that recently
obtained for the massless case, and it is the
two-dimensional version of the low-energy QCD chiral Lagrangian at
finite baryon density.

\end{abstract}

\narrowtext

\section{Introduction}\dle{sec:intro}

The sine-Gordon (SG) theory in two dimensional Minkowski space-time
with metric $(+,-)$ is described by the Lagrangian density
\be
{\cal L}_{SG}[\phi] = \frac{1}{2} \paa_\mu \phi  \paa^\mu \phi +
\frac{\alpha_0}{\lambda^2} \cos \lambda \phi - \gamma_0,
\dle{SG}
\ee
where $\phi$ is a real scalar field, $\gamma_0$, $\alpha_0$ and
$\lambda$ are bare parameters to be renormalised later, and $x^{\mu} =
(t,x)$, $\mu = 0,1$.  Notice that apart from the usual kinetic term,
the potential term is periodic so that there are an infinite number of
degenerate vacua whose value, $\phi_v$, depends on the coupling
constant
$\lambda$: $\phi_v = 2 n \pi/\lambda$ where $n \in \IZ$.  
We comment that the Lagrangian (\dr{SG}) is invariant under
$\phi \rightarrow \phi + \phi_v$ and $\phi \rightarrow -\phi$
meaning that the sign of $\alpha_0$ is unimportant, and the name
sine-Gordon is related to the fact that the equations of motion
contain a $\sin \phi$ term (see \dc{colebook,Raj}).

The massive Thirring (MT) model, on the other hand, is a model with a
fermionic field $\psi = (\psi_1, \psi_2)$ and a 
four Fermi interaction with coupling $g$:
\be   
 {\cal{L}}_{Th} [\bar\psi,\psi] =  i\bar{\psi} (\pabar - m_0)\psi +
 \frac{1}{2}g  j_\mu (x) j^\mu (x) , 
\dle{Th}
 \ee    
where $\curr=\bar\psi (x)
\gamma_{\mu} \psi (x)$ and 
$m_0$ is a bare mass.  For positive coupling constant, 
$g >0$, the interaction term is attractive and there are
fermion-antifermion bound states.

Though it maybe somewhat surprising since one is
a model with a bosonic field and the other one with a fermionic field,
it is
very well known that these two theories are linked.  In
particular, they  provide one of the earliest examples of 
duality in which the weak limit of one theory describes the same
physics as the strong limit of the other and conversely since
\be
\frac{\lambda^2}{4\pi}= \frac{1}{1+g/\pi}.
\dle{equivconst1}
\ee
Hence perturbative calculations in one theory tell us about
non-perturbative effects in the other.  Identity
(\dr{equivconst1}) was proved in perturbation theory about
$\alpha_0 = 0 = m_0$ (see below) at $T= \mu = 0$ by Coleman using
canonical operator methods
\dc{Coleman} and later on using path integrals \dc{fugasc82,na85}. 
Note the special value $g=0$ in which the Thirring model is a free
theory of massive fermions, corresponding to $\lambda^2 = 4
\pi$ in the SG model.  Also the MT bound states exist for
$g > 0 \Rightarrow \lambda^2 < 4 \pi$.  

The duality (\dr{equivconst1}) is directly related to the fact that
the models exhibit bosonisation, in which a theory of fermions is
equivalent to a theory of bosons.  In general the bosonic (fermionic)
theory may also have fermionic (bosonic) excitations;  this indeed
occurs in the SG and MT models and we will see another example below
in low-energy QCD.  In the case of the SG and MT models, bosonisation
schematically works in the following way (see
\dc{Raj} for a summary as well as a list of the relevant articles). 
Consider first the SG model.  Perturbations in $\lambda$ about one of
the minima of the degenerate potential give rise to the usual
bosonic simple harmonic oscillator
spectrum.  The model also has other excitations.  Recall that
there are topologically non-trivial
solitonic
solutions to the classical equations of motion,
and that these arise from the vacuum degeneracy:  there
is no reason why at $x
\rightarrow
\infty$ the system should be in the same vacuum as at $x \rightarrow
-\infty$.  As a result one can have finite energy solutions with
non-zero charge
\be
Q = \int_{-\infty}^{\infty} dx \frac{\lambda}{2 \pi} \frac{\paa}{\paa
x^1}\phi(x) = \frac{\lambda}{2 \pi} \left[ \phi(\infty) -
\phi(-\infty) \right]  = \Delta n.
\dle{Qdef}
\ee
The simplest `kink' solution has $Q=1$ and an energy proportional to
$\alpha_0/\lambda^2$, whilst the anti-kink ($Q=-1$) is obtained from
the kink by taking $\phi \rightarrow - \phi$.  Furthermore, since the
system is integrable, exact expressions for multi-kink and anti-kink
states are known and one can construct moving kinks by Lorentz
transformation.  These may then be quantised using semi-classical
methods \dc{Raj,Dashen}.  There are also classical solutions with
$Q=0$ corresponding to `breather' solutions in which kinks and
anti-kinks oscillate periodically about each other \dc{Raj}.  These can
be quantised by WKB methods \dc{Raj,Dashen}.  The relationship between
these SG excitations and the fundamental fermionic excitations and
bound state excitations of the MT model is summarised schematically in
table I.
\typeout{--- sgth table number 1 ---}

\begin{center}
\begin{tabular}{|p{3cm}|p{5cm}|}
\hline \hline
MT MODEL & SG MODEL \\
\hline \hline 
fermionic fields & bosonic fields \\
\hline
$g$ & $\lambda \sim 1/g$ \\
\hline
 & \\
$\bar{f} f$ bound states ($g>0$)& 
bosons, kink -- anti-kink 
breather solutions ($\lambda^2 < 4 \pi$) \\ 
\hline
$f$ & kinks \\
\hline
+ $\mu j^0$  & $m_0 \neq 0 \rightarrow$ see later 

 $m_0 = 0 \rightarrow$ topological term \dc{AlvGom98} \\
\hline \hline
\end{tabular}
\vskip .3cm

 TABLE I.  
Table showing schematically some of the links between the SG
and MT models at $T = 0$
\end{center}
%
%
We see that the  kinks themselves are fermion-like
excitations corresponding to the fundamental fermions of the Thirring
model.  For
$\lambda^2 < 4 \pi$, the bosonic ($Q=0$) excitations of the SG model
correspond to the bound states of the MT model.  This is what is meant
by bosonisation.

What is the motivation for studying these models at $T > 0$ and $\mu
\neq 0$?  One first reason is that these
models could be used to develop techniques which may then be applicable
to more realistic 4D theories, in particular QCD.  Recall that at low
temperatures, where the quarks and gluons are strongly confined into
hadrons, one can describe the system by an effective chiral bosonic
lagrangian (CBL) for the lightest mesons (pions, kaons and eta) which are the
Nambu Goldstone bosons (NGB) of the chiral symmetry breaking.  The
relationship between this chiral bosonic theory and the original
fermionic QCD has many similarities with the relationship between the
SG and MT models (see table II).
\typeout{--- sgth table number 1 ---}
\renewcommand{\arraystretch}{1}
\begin{center}
\begin{tabular}{|p{3cm}|p{5cm}|}
\hline \hline
QCD & CBL \\
\hline \hline
quarks and gluons& bosonic fields \\
\hline
$\alpha_s$ (large at low energies) & small expansion parameter
$\frac{m_{\pi}}{\Lambda_{\chi}}$, $\Lambda_{\chi}\simeq 4\pi f_\pi 
 \simeq$ 1 GeV
\\
\hline
 & \\
$ \bar{q}q$ bound states   & $\pi$, $K$, $\eta$ (NGB of chiral symmetry)\\
\hline
 baryon & skyrmion (topological defect) \\
\hline
+ $\mu$ for baryons & topological term \dc{AlvGom95} \\  
\hline \hline
\end{tabular}
\vskip .3cm

TABLE II. Table showing schematically some of the links between QCD and
the low energy chiral lagrangian.
\end{center}
For example, the
bound states of QCD are the mesons of the chiral theory whilst
the baryons of QCD correspond to skyrmions---topological defects
\dc{sk61,witt83}---in the chiral theory which to lowest order is the
non-linear sigma model \dc{Angbook} 
({\em c.f}.\ the relationship between the SG
and MT models in table I). 
Note also that the SG Lagrangian in (\dr{SG}) corresponds
to a non-linear sigma model in 2D for a single Nambu-Goldstone-like
field
$\phi$. Although there is no spontaneous symmetry breaking in 2D
\dc{col73}, the potential term in (\dr{SG}) breaks explicitly
the symmetry $\phi\rightarrow\phi+a$  with $a\in\IR$ (which we 
will call the chiral symmetry for reasons to become clear below) 
still preserving the symmetry  $\phi\rightarrow\phi+ \phi_v$. These two
symmetries are,  respectively, the counterparts of the  chiral and
isospin symmetries for QCD, $\alpha_0$ and $\lambda$ playing the r\^ole
of the pion mass squared and the inverse of the pion decay constant $f_\pi$ 
respectively.  On the other hand, the chiral symmetry transformations
in terms  of the Thirring fermion are
$\psi\rightarrow\exp(ia\gamma_5)\psi$. The massless Thirring model is
chiral invariant, the fermion  mass term breaking that symmetry in the
same way as the $\alpha$ term does in the SG Lagrangian.

A second motivation for studying these models at $T>0$ and $\mu \neq
0$ comes from cosmology.  The SG kinks are the 2D analogues of cosmic
strings, line-like defects formed in 4D when a system goes through a
symmetry breaking phase transition (say of some group $G$ to a
subgroup $H$) for which the first homotopy group $\pi_1(G/H) \neq 0$
\dc{Kib76,ViSh}.  The strings trap regions of the unbroken high energy
phase and so have energies per unit length which can be very large,
depending on the critical temperature.  In the context of the early
universe and cosmology, cosmic strings may have played an important
r\^{o}le because as the universe cooled and expanded after the big
bang, it went through a number of phase transitions some of which may
have led to the formation of strings.  In particular, strings formed
at the GUT phase transition have huge energies per unit length
($\mu \sim 10^{32}$ GeV$^2$) and hence significant
gravitational effects, and so it is thought that they may have been
responsible for the temperature fluctuations in the cosmic microwave
background radiation and for seeding the perturbations which led to
the formation of structures such as galaxies \dc{ViSh}.

However, in order to make detailed predictions as to their
effects, it is important to know the initial distribution of strings and
whether or not it contains infinite strings.  This is a very difficult
task to undertake analytically \dc{Ray}, but recently progress has
been made by using the analogy between experimentally observable
systems such as $^3$He and $^4$He and the early universe
\dc{GKHe3}.  It would seem, however, that there is an even more simple
experimentally accessible system with which one could try to test
ideas of defect formation, and that is a Josephson junction
\dc{JJbook}.   This device consists of two layers of superconductors
separated by a thin  dielectric barrier, typically of the order of
5nm.  Denoting the
macroscopic wave function of each of the superconductors by $\Psi_a =
|\Psi_a|\exp \{i \theta_a \}$ where $a = 1,2$ labels each of the two
layers, then Josephson tunnelling of the Cooper pairs across the
dielectric layer results in $\Delta \theta = \theta_1 -
\theta_2$ satisfying a SG equation \dc{JJbook}.  As the two
superconductors are taken through the phase transition, kinks
are formed in the junction and these are observed experimentally
\dc{Roberto1}.  Experiments are also done to see what is, for example,
the effect of the geometry of the set up on the kinks and their
dynamics \dc{Roberto2} and these devices are used as sources of some of
the highest energy micro-waves finding applications in satellites
\dc{JJbook,Roberto1}.   One idea might therefore be to see whether one
can indeed use such experiments with Josephson junctions to test ideas
of defect formation.  The hope is in particular that the situation can
be much simplified in this case because of the duality between the SG
and MT models.  First, however, one has to check what is the relation
between these two models at $T>0$ and
$\mu \neq 0$.  In particular, could it be possible that the
relationship between coupling constants (\dr{equivconst1}) is a
function of temperature?  As we will comment in the conclusions, these
models also provide examples of other phenomena which may too be more
easy to study in 2D rather than 4D.

Here we report on our first steps in these directions.  In section
\dr{sec:res2D} we briefly summarise some of the particularities and
basic results of 2D which will be useful to bear in mind for the rest
of the work.  Results are then given in section \dr{sec:res}. 
First, in subsection \dr{sec:op}, we outline the main steps which must
be taken to extend Coleman's work \dc{Coleman} on the SG model 
 to $T>0$ using an operator approach.  The results
of \dc{Belg} showing the equivalence of the partition functions are
reproduced though the approach is entirely different. Path integral
methods are used in the rest of the work.  We show that not only the
partition functions are equivalent but also that thermal averages of
correlators of  zero-charge operators evaluated at different
space-time points coincide.  The relationship between normal ordering
and regularisation is specified and we analyse the $T>0$ situation in
which there is a net number of fermions,
$\mu\neq 0$.  The partition function is calculated in that case so
extending \cite{AlvGom98} to the massive case and \cite{Belg} to
$\mu\neq 0$.  The analogy between these models and a classical gas of
particles is then noted in the conclusions.  Details of the
calculations and results presented here may be found in
references \dc{GS,GRS}.

\section{Peculiarities of 2D}\dle{sec:res2D}

The following three basic, though important, results hold for free
fields in 2D at $T>0$.  They will be useful in the following
sections.  
\\
\\
{\bf 1)  Equality of free massless boson and fermion partition
functions}
\\
Consider a free bosonic field of mass $\tilde{\mu}$ and a massless
free fermionic field.  The partition functions can be obtained by
writing the Hamiltonians in terms of a normal ordered part plus an
infinite vacuum energy; they are respectively
\ba
\ln Z_0^{\tilde{\mu},B}(T) & = &
 -L\intk\left[\frac{\beta\omega_{k,\tilde{\mu}}}{2}+ \right.
\nonumber
\\
&  &  \; \; \; \; \; \; \; \; \; \; \; \;+ \left.
\ln\left(1-e^{-\beta\omega_{k,\tilde{\mu}}}\right)\right],
\dla{partbos}
\\
\ln Z_0^{F}(T) &=& 2 L \intk \left[\frac{\beta \vert k\vert}{2}
\right.
\nonumber
\\
& &
\; \; \; \; \; \; \; \; \; \; \; \;+ \left. \ln \left( 1+e^{-\beta
\vert k\vert}
\right)
\right].
\dla{partferm}
\ea
Here $\omega_{k,\tilde{\mu}} = k^2 + \tilde{\mu}^2$, $\beta$ is the
inverse temperature $\beta = 1/T$, and
$L$ is the spatial dimension which we are taking to infinity.  As
usual, all the thermodynamic observables are obtained from the
logarithm of the 
partition function and its derivatives by dividing by $\beta L$.

Observe that when $\tilde{\mu}=0$, 
\ba
\lefteqn{ \intk \ln\left(1-e^{-\beta \vert k\vert}\right) }
\nonumber
\\
&=&
-2\intk \ln\left(1+e^{-\beta \vert k\vert}\right)=-\frac{\pi T}{6},
\nonumber
\ea
and hence it follows that ignoring vacuum terms, the two partition
functions (\dr{partbos})-(\dr{partferm}) are equivalent for
$\tilde{\mu}=0$:
\be
Z_0^{F}(T) = Z_0^{0,B}(T) = \exp {\left[\frac{\pi L
T}{6}\right]} .
\dle{pffree}
\ee
Thus one cannot tell the difference between the bulk quantities of
these two systems:  equality (\dr{pffree}) is perhaps the simplest
example of bosonisation.  As will be seen below, the equality of the
SG and MT partition functions at $T>0$ rests on (\dr{pffree})
since in those models we work in perturbation theory, expanding about
$\alpha_0 = 0$ in the SG model and about $m_0=0$ in the MT model
(and hence about massless bosonic and fermionic theories).
\\
\\
{\bf 2)  Free thermal propagators}
\\
We will need the free propagators for boson and fermion fields
at $T>0$.  Again working with a boson field of mass $\tilde{\mu}$ and a
massless fermion field, and calculating in the
imaginary time formalism with  $\bar{t} = i t$, these are
found to be respectively \dc{Belg}
\ba
\Delta_T(x) &=&-\frac{1}{4\pi}\ln \tilde{\mu}^2\beta^2 Q^2(x)+ K +
\Od(\tilde{\mu}\beta),
\dla{bosprop}
\\
S_{\alpha\beta} (x)&=&
-\frac{1}{2\beta}\frac{\Qbar_{\alpha\beta} (x)}{Q^2(x)},
\dle{fermprop}
\ea
where we have expanded the boson propagator about $\tilde{\mu} =
0$, and $K$ is a constant.  The $Q$ variable is given by
$Q^2=Q_0^2+Q_1^2$ where
\ba
Q_0(x,\bar{t})&=&-\cosh (\frac{\pi x}{\beta}) \sin (\frac{\pi
\bar{t}}{\beta}),
\nonumber
\\ 
Q_1(x,\bar{t})&=&-\sinh (\frac{\pi x}{\beta}) \cos
(\frac{\pi \bar{t}}{\beta}),
\dla{qs}
\ea
so that $Q(x)$ is a Lorentz scalar. 
The indices $\alpha$, $\beta$ are Dirac indices and we have worked
with the 2D Euclidean $\gamma$ matrices
$$
\gamma^0=
\left( \begin{array}{cc} 
   0 & 1 \\   
  1    &   0  
\end{array} \right) \,  \, \, \,    
\gamma^1=\left( \begin{array}{cc}     
    0 & -i \\      
       i    &   0  
\end{array} \right)
$$
$$
\gamma_5 = -i \gamma^0 \gamma^1 =
\left( \begin{array}{cc}    
          1 & 0 \\        
    0    &   -1    
\end{array} \right)     
$$
so that
$$
\{\gamma^\mu,\gamma^\nu\} = 2\delta^{\mu\nu},
\; \; \; \; \; 
\gamma_\mu\gamma_5 = -i\epsilon_{\mu\nu}\gamma_\nu 
$$
with $\epsilon_{01}=1$.  Observe that (\dr{bosprop}) is both 
ultra-violet divergent ($x \rightarrow 0$) as well as infra-red
divergent ($x \rightarrow \infty$ or $\tilde{\mu} \rightarrow 0$) since
\ba
Q^2(x,\bar{t})&\stackrel{\vert x\vert\rightarrow\infty}
{\longrightarrow}&\frac{1}{4}
e^{ 2\pi \vert x\vert /\beta} \qquad \forall \bar{t}\nonumber\\
&\stackrel{(x,\bar{t})\rightarrow (0,0)}{\longrightarrow}&(\pi T)^2
(x^2+\bar{t}^2).
\dla{asbeq}
\ea
An interesting property of the propagators
(\dr{bosprop})-(\dr{fermprop}) is that they can be directly obtained
from the corresponding $T=0$ propagators through the
substitution $x_{\mu} \rightarrow \beta Q_{\mu}$.  This is in fact
true for all contours in the complex time plane, and can be viewed
either as a result of the conformal invariance of the free theories,
or as arising from solving the Green function equation for the Coulomb
potential on a cylinder of radius $\beta$ \dc{GRS}.
\\
\\
{\bf 3)  Ultra-violet divergences and normal ordering}
\\
Finally, recall that the UV divergence structure of 2D bosonic
theories is much simpler than that of 4D ones.  The reason follows from
the fact that in $d$ dimensions with an interaction of the form
$\phi^r$, a diagram with $n$ vertices and $E$ external lines has a UV
degree of divergence $D$ of \dc{Ryder}
$$
D=d - \left( \frac{d}{2}-1 \right)E + n \left[ \frac{r}{2} (d-2) -d
\right].
$$
So with $d=2$,
$$
D = 2 - 2n
$$
and the only divergent diagrams $\forall r$ are tadpole diagrams.  
Since the SG lagrangian contains a term $\cos \lambda \phi = \sum
(-1)^2 \phi^{2n}/(2n!)$ then in  the operator formalism it
follows that all UV divergences of the theory should be removed
through normal ordering (as this removes tadpole
diagrams).  In path integral methods there is no
operator normal ordering, and the divergences are removed using
different methods (see below).

\section{Results}\dle{sec:res}

\subsection{Bosonisation in the canonical operator approach at $T>0$
and
$\mu=0$}\dle{sec:op}

We have used operator methods to extend the paper of Coleman
\dc{Coleman} to $T>0$ and $\mu = 0$.  This enabled us to prove
\dc{GS} that the partition functions of the SG and MT models are
identical
\be
Z_{SG}(T,\mu = 0) = Z_{MT}(T,\mu = 0)
\dle{identmuzero}
\ee
provided the coupling constants of the theories satisfy
(\dr{equivconst1}) which is  temperature independent.  In
fact, (\dr{identmuzero}) was already proved using very different path
integral methods in
\dc{Belg} (see also section \dr{sec:pi1}).

In the SG model the main steps in the calculation $Z_{SG}(T,\mu=0)$
are the following:
\begin{enumerate}
\item  Remove all UV divergences by normal ordering the SG Hamiltonian.
\item  Deal with the IR divergence of the propagator (\dr{bosprop}).
\item  Calculate the partition function $Z_{SG}$, a sum of thermal
expectation values of free (interaction picture) operators each of
which have been normal ordered as a result of step 1.
\end{enumerate}
We now outline the main features of each of these steps, and also
comment on the differences between this $T>0$ case and the
$T=0$ one discussed in \dc{Coleman}.
\\
\\
{\bf Step 1:  Removal of all UV divergences by normal ordering}
\\
As stated in point 3 above, the partition function contains thermal
expectation values (TEV's) of normal ordered operators.  Therefore to
simplify its calculation, we remove UV divergences by using thermal
normal ordering (TNO) introduced in \dc{ES} rather than standard zero
temperature normal ordering (which places annihilation operators to
the left of creation operators). By construction TNO, which is
denoted by $N^{ES}$, guarantees that for any\footnote{This could be
the field operator $\hat{\phi}$,  the momentum operator
$\hat{\pi}$ or any other  composite operator.} operator  
(other than the identity) $\bullet$  in the interaction picture,
\be
\ll N^{ES} [\bullet] \gg_0 \; \;  = 0,
\dle{NO}
\ee
whereas for usual normal ordered products
$$
\ll 
:\bullet: \gg_0  \; \; \neq 0.
$$ 
The operation $N^{ES} [\bullet]$ was defined in \dc{ES} to place the
``positive'' parts of the operator, $\bullet^+$ (a combination {\em
both} of annihilation {\em and} creation operators), to the right of
the ``negative'' part,
${\bullet}^-$.  See \dc{ES} for the exact definitions.  
In the above equations the angular brackets denote a thermal
expectation value, and the subscript zero indicates that we work with
free (interaction picture) fields, and hence in perturbation theory. 
Identity (\dr{NO}) will greatly simplify calculations below.  

The SG Hamiltonian is
\ba
\hat{H}_{SG} &=& \int_0^{L} dx \left[ \frac{\hat{\pi}^2}{2} +
\frac{1}{2}
\left(
\frac{\paa
\hat{\phi}}{\paa x} \right)^2 -
\frac{\alpha_0}{\lambda^2} \cos \lambda \hat{\phi} - \gamma_0 \right]
\nonumber
\\
&=:&   \hat{H}_0 - \int_0^{L} dx  \left(
\frac{\alpha_0}{\lambda^2} \cos \lambda \hat{\phi} + \gamma_0 \right)
\dla{HSG}
\ea
which must be divided into a free and interacting part so as to apply
TNO.  Although the term $\cos \lambda \hat{\phi}$ itself contains a
mass term on expansion in powers of $\lambda$, we want to
keep $\lambda$ of arbitrary size.  Consider therefore
\ba
\hat{H}_{SG} &=& 
\left[ \hat{H}_0  + \int_0^{L} dx  \left( \half \rho^2 \hat{\phi}^2 \right)
\right]
\nonumber
\\
&-& \left[  \int_0^{L} dx \left(
\frac{\alpha_0}{\lambda^2} \cos \lambda \hat{\phi} + \half \rho^2
\hat{\phi}^2 + \gamma_0  \right)
\right],
\nonumber
\ea
so that perturbations are about a scalar field of mass
$\rho$.  To take account of this fact, TNO is now denoted by
$N^{ES}_{\rho} [\bullet]$ and similarly we add an extra mass label to 
 the propagators (\dr{bosprop});  $\Delta_{T}(x) \rightarrow
\Delta_T(x;\tilde{\mu})$.  Hence in the case of $\hat{H}_{SG}$ above
we will be dealing with $\Delta_T(x;\rho)$.  The link between $\rho$
and the regularisation scale in path-integral methods is
discussed in section \dr{sec:pi1}.

TNO of (\dr{HSG}) may be carried out by using the identity
\ba
e^{i \int_c d^2x j(x) \hat{\phi}(x)} &=& N^{ES}_{\rho} 
\left[ e^{i \int_c d^2x j(x)
\hat{\phi}(x)} \right] \times
\nonumber
\\
&& 
\; \;    e^{\frac{1}{2} \int_c d^2x 
\int_c d^2y j(x)
\Delta_T(x-y;\rho) j(y)}
\nonumber
\\
&=& 
T_c \left[ e^{i \int_c d^2x
j(x) \hat{\phi}(x)} \right]
\dla{identity1}
\ea
where $T_c$ means contour ordering\footnote{These equalities may at
first sight seem surprising, but they might be clarified by noting
that here the advanced and retarded thermal propagators
are equal so that $\Delta^>_T(x-y,\rho) =
\Delta_T(x-y,\rho) = [\phi^+(x),\phi^-(y)]$ where the positive and
negative parts refer to those of \dc{ES}.}.  Had we used
$T=0$ normal ordering, the zero-temperature propagator would have
appeared in (\dr{identity1}) rather than the
finite temperature one:  TNO means that the $Q$ variables of
(\dr{qs}) are built in from the start of the calculation.  
Finally the UV divergence of
$\Delta_T(x;\rho)$ is regulated by cutting off the theory  replacing
\be
\Delta_T(x;\rho) \; \; \; \longrightarrow \; \; \;
\Delta_{T}(x;\rho;\Lambda) :=
\Delta_{T}(x;\rho) -
\Delta_{T}(x;\Lambda)
\dle{regprop1}
\ee
where $\Lambda$ is a large mass. 
Hence the constant $K$ in the propagator (\dr{bosprop}) 
cancels, and $\Delta_{T}(x;\rho;\Lambda)$ is now both
non-singular as well as $\beta$ independent for $x \rightarrow 0$; 
$
\Delta_{T}(0;\rho;\Lambda)  = -\frac{1}{4\pi} \ln 
\left( {\rho^2
}/{\Lambda^2} \right).
$
Combining (\dr{regprop1}) with (\dr{identity1}) for $j(x) =
\lambda
\delta (x-y)$ gives, for example
\be
e^{i \lambda \hat{\phi}(y)} = \left( \frac{\rho^2
}{\Lambda^2} \right)^{\frac{\lambda^2}{8\pi} } N^{ES}_{\rho} 
\left[ e^{i
\lambda \hat{\phi}(y)} \right].
\dle{noexpphi}
\ee
After similar manipulations, (\dr{HSG}) can be written as \dc{GS}
\be
\hat{H}_{SG} =  N^{ES}_{\rho} \left[ \hat{H}_0 -
\frac{\alpha}{\lambda^2}\int_0^{L} dx
\cos \lambda \hat{\phi}  - L \gamma \right]
\dle{HSGren}
\ee
where $\alpha_0$ has been multiplicatively renormalised,
\be
\alpha = \alpha_0  \left( \frac{\rho^2
}{\Lambda^2} \right)^{\frac{\lambda^2}{8\pi} }  
\dle{alpren1}
\ee
and $\gamma_0$ has been renormalised according to
\be
\gamma = \gamma_0 - E_T(\rho)
\dle{gam}
\ee
where 
\ba
E_T(\rho) &=& \frac{1}{2} \left\{ \left[\hat{\pi}^+,\hat{\pi}^-
\right] + \left[(\paa_0 \hat{\phi})^+,(\paa_0 \hat{\phi})^-
\right] \right\}
\nonumber
\\
&=&
 E_0(\rho) +   L \intk  \frac{1}{2}
N_{k,\rho}
\left( \frac{2k^2 +
\rho^2}{\omega_{k,\rho}} \right).
\nonumber
\ea
Here $E_0(\rho)$ is an infinite temperature independent contribution;
$E_0(\rho) =   L \intk  \frac{1}{4} \frac{2k^2 +\rho^2}
{\omega_{k,\rho}}$.  The temperature dependent part 
proportional to the Bose-Einstein distribution  $N_{k,\rho}= (e^{\beta
\omega_{k,\rho}} - 1)^{-1}$ is finite.
The coupling $\lambda$ is unchanged.

Thermal normal ordering the SG Hamiltonian has therefore absorbed
all UV
infinities just as zero temperature normal ordering does \dc{Coleman},
but it has also introduced some extra $T$-dependent finite terms.  
\\
\\
{\bf Steps 2 and 3:  IR divergences and calculation of the 
 partition functions}
\\
The IR divergence of the boson propagator
(\dr{bosprop}) is removed by introducing a mass $\tilde{\mu}$ into the
SG Hamiltonian (\dr{HSGren}).  At the end of the calculation we let
$\tilde{\mu}
\rightarrow 0$, and hence are free to add the
extra mass term within the normal ordering giving the
Hamiltonian
\ba
\hat{H} &=&  N^{ES}_{\rho} \left[ \left( \hat{H}_0^{\tilde{\mu}} -
\gamma L \right)
 -
\frac{\alpha}{\lambda^2} 
\int_0^L dx \left( \cos \lambda \hat{\phi} \right)   \right]
\nonumber
\\
&=& N^{ES}_{\rho} \left[ \hat{{\cal{H}}}_0^{\tilde{\mu}}
 -
\frac{\alpha}{\lambda^2}\int_0^L dx \left(  \cos \lambda \hat{\phi}
\right)  
 \right]
\nonumber
\\
&=:& \hat{A}_0 + \hat{A}_{I}.
\ea
Here $\hat{H}_0^{\tilde{\mu}} = \hat{H}_0 + \int_0^L dx 
\frac{1}{2}
\tilde{\mu}^2
\hat{\phi}^2 $, and
$ \hat{A}_0$ and $\hat{A}_{I}$ denote
the free ($\alpha=0$)  and interacting
Hamiltonians respectively.
In perturbation theory the SG partition function is therefore given by
\ba
Z_{SG}(T,\mu = 0) &=& \lim_{\tilde{\mu} \rightarrow 0} \Tr 
\left\{e^{-\beta \hat{H}}
\right\} 
\nonumber
\\
&=&
\lim_{\tilde{\mu} \rightarrow 0}
\sum_n 
\frac{1}{n!} 
\int_{0}^{\beta} dt_1 \ldots \int_{0}^{ \beta} dt_n  \times
\nonumber
\\
&& Tr \left\{
e^{-\beta \hat{A}_0} T_c \left[ \hat{A}_I(t_1) \ldots  \hat{A}_I(t_n)
\right]
\right\},
\dla{partz}
\ea
so that one only needs calculate free expectation values
$ \ll \bullet \gg_0  = \Tr \exp \{-\beta \hat{A}_0 \bullet \} /
\Tr \exp \{-\beta \hat{A}_0 \}$.\footnote{In fact one has to be
extremely careful when considering the precise form of the thermal
weight
$\exp \{ -\beta \hat{A}_0 \}$.  The reason is that it contains two {\em
different} mass scales $\rho$ and $\tilde{\mu}$, meaning that 
 $\hat{A}_0$ is
{\em not} obviously  diagonal and so not obviously of the form $\int
dk \,
\omega(k) \hat{a}^{\dagger}(k) \hat{a}(k)$ as is usually the case and
as was assumed in \dc{ES}.  See
\dc{GS} for details about this point.}
Indeed, the correlator appearing in  (\dr{partz}) can be obtained by
showing first that \dc{GS}
\ba
\lefteqn{ \ll T_c \left[ \prod_{j=1}^{n} N_{\rho}^{ES} 
\left[ e^{i  \lambda_j \hat{\phi}(x_j)}\right] \right] \gg_0}
\nonumber
\\
&=& 
\left( \frac{\tilde{\mu}^2}{\rho^2}
\right)^{\frac{\sum_j \lambda_j^2}{8\pi}}\prodj  
\ll N^{ES}_{\tilde{\mu}} \left[ e^{i 
\sum_j (\lambda_j \hat{\phi}(x_j))} \right] \gg_0 \times
\nonumber
\\
&&\; \; \; \; \; \; \; \; \; \; 
\times \prod_{j>k}^n \left[ \beta^2 
\tilde{\mu}^2 |Q(x_j - x_k)|^2 \right]^{\frac{\lambda_j
\lambda_k}{4 \pi}}
\nonumber
\\
& = & 
\left( \frac{\tilde{\mu}^2}{\rho^2}
\right)^{\frac{1}{8\pi}\sum_j \lambda_j^2}  \prodj\prod_{j>k}^n \left[
\beta^2
\tilde{\mu}^2 |Q(x_j - x_k)|^2 \right]^{\frac{\lambda_j
\lambda_k}{4 \pi}},
\dla{answer}
\ea
where we have expanded the exponential inside the
normal ordered term and noted that all terms which contain
$\hat{\phi}^n$ vanish (by (\dr{NO})) apart from that with 
$n=0$.  Then observe that the terms
proportional to $\tilde{\mu}$ in (\dr{answer}) have a
contribution $\tilde{\mu}^{{\left(\sum\lambda_j \right)^2}/{4
\pi}}$.  Thus in the limit $\tilde{\mu} \rightarrow 0$, only
configurations with $\sum
\lambda_j = 0$ contribute (a condition which will become analogous to
the fermion chiral selection rule of section \dr{sec:pi1}). In
(\dr{answer}) $n$ must therefore be even.  If we let $m=n/2$, 
$y_j=x_j$ for $j=n/2+1,\dots,n$ and define
\be
\hat{A}_{\pm} = N^{ES}_{\rho} \left[ e^{\pm i
\lambda \hat{\phi}} \right],
\dle{Adef}
\ee
then from (\dr{answer})
\ba
\lefteqn{ \ll T_c \prod_{j=1}^{m}  \hat{A}_+(x_j)\hat{A}_-(y_j) \gg_0
\; = }
\nonumber
\\
&&\prod_{j=1}^{m}
\frac{ \prod_{j > k}^{m} \left[ \beta^4
\rho^4 |Q(x_j - x_k)|^2 |Q(y_j - y_k)|^2
\right]^{\frac{\lambda^2}{4\pi}} }{
 \prod_{k=1}^m \left[ \beta^2
\rho^2 |Q(x_j - y_k)|^2 \right]^{\frac{\lambda^2}{4\pi}} }
\dla{Acorrel}.
\ea
Hence the partition
function (\dr{partz}) is \dc{GS}
\ba
\lefteqn{ Z_{SG}(T,\mu = 0) = }
\nonumber
\\
&& 
Z_0^{B}(T)
 \sum_n  \left( \frac{1}{(n!)} \right)^2  \left[
\frac{\alpha}{2\lambda^2}
\left( \frac{T}{\rho} \right)^{\frac{\lambda^2}{4 \pi}} \right]^{2n} 
\prod_{j=1}^n \int_{T} d^2x_j  d^2y_j 
\times
\nonumber
\\
&&
\; \; \; \; \; 
\frac{ \prod_{j > k}^{n} \left[ |Q(x_j - x_k)|^2 |Q(y_j - y_k)|^2
\right]^{\frac{\lambda^2}{4\pi}} }{ 
 \prod_{k=1}^n \left[  |Q(x_j - y_k)|^2
\right]^{\frac{\lambda^2}{4\pi}} }
\dla{sgparfun}
\ea
where $\intT\equiv \int_0^\beta dt\int_{-\infty}^{+\infty}
dx$ and $Z_0^{B}(T) =  \Tr \left\{   e^{-\beta \hat{A}_0} \right\} = e^{\beta
\gamma_0 L } Z_0^{0,B}(T)$ which is finite for $\gamma_0$ satisfying
(\dr{gam}) and $\lambda^2 < 4 \pi$.\footnote{From the behaviour 
of the $Q$ variables in (\dr{asbeq}), one
can see that for
$\lambda^2 > 4 \pi$ there are extra divergences in (\dr{sgparfun}).
The treatment of these is commented on in \dc{GS}.}

For the MT model, calculation of $Z_{MT}(T,\mu = 0)$ in the operator
approach is rather more complicated (it is based on a paper by
Klaiber \dc{Klaib}) and is discussed in
\dc{GS}.   In perturbation theory one can show that the partition function
$Z_{MT}(T,\mu = 0)$ is given by \dc{GS} 
\ba
\lefteqn{ Z_{MT} (T,\mu = 0)= }
\nonumber
\\
&&
Z_0^F (T)\sumn \left(\frac{1}{n!}\right)^2 
\left[\frac{m}{2\beta}\left(\frac{\rho}
{T}\right)^{\kappa^2/\pi}\right]^{2n} \prodjint \times
\nonumber
\\
&& \; \; \; \; \;
\frac{\prod_{k<j}\left[
Q^2(x_j-x_k)Q^2(y_j-y_k)\right]^{1-\kappa^2/\pi}}{
\prod_{k=1}^n\left[Q^2(x_j-y_k)\right]^{1-\kappa^2/\pi}}
\dla{thparfun}
\ea
where we have chosen to renormalise the MT model at the scale
$\rho$.  Here
$Z_0^F$ is given in (\dr{partferm}), 
\be
\kappa^2 = \frac{g}{1+g/\pi}
\dle{kapdef}
\ee
and $m$ is the renormalised mass (see \dc{GS} and also section
\dr{sec:pi1})
\be
m = m_0 (\Lambda/\rho)^{\kappa^2/\pi}.
\dle{mren}
\ee
Thus term by term (\dr{sgparfun}) and (\dr{thparfun}) are
identical provided that a) the parameters of the two theories are
identified as in (\dr{equivconst1}), b) that $\alpha$
and $m$ are related by
\be
\rho m = \frac{ \alpha}{\lambda^2}
\ee 
and c) that $\gamma_0 = 0$ as then $Z_0^{B}(T) = Z_0^F(T)$ by
(\dr{pffree}).

We have therefore extended the work of Coleman
\dc{Coleman} to $T>0$, $\mu = 0$ using operator methods.  This was
rendered more simple through the use of TNO.  The results obtained are
in fact the same as those of
\dc{Belg} though the approach has been entirely different.  We now
turn to path integral methods.

\subsection{Path Integral bosonisation at $T>0$ and $\mu =
0$}\dle{sec:pi1}

We have used path integral methods not only to prove the equality of
the two partition functions as in (\dr{identmuzero}) 
\dc{GS,Belg}, but also to prove the equivalence of certain sets of
correlators of operators evaluated at {\em different} space time
points.  These correlators cannot be obtained from the partition
function which only contains information about global thermodynamic 
observables like the pressure or the condensates, but not about 
correlators which physically yield, for instance, thermal correlation
lengths.  We outline the main points of such PI calculations for
the SG model and then the MT model.  This will enable the link between
normal ordering and PI regularisation to be made.  Our results for
the correlators will also be stated more precisely.

\subsubsection{SG model}

As always in path integral methods, one works with the
generating functional. Once again it is useful to start with free
boson fields and in particular to calculate the correlator
(\dr{Acorrel}) which may be obtained from the free boson Euclidean
generating functional
\ba
Z_0^B[J;T]&=&N_\beta
 \pib \exp \left\{ -\left[ \intT \frac{1}{2} \right. \right. \times
\nonumber
\\
&&
\; \; \; \; \; \left. \left. 
\left[
(\partial_\alpha \phi)^2+
{\tilde{\mu}}^2\phi^2\right]+J(x)\phi(x)\right]\right\}
\nonumber
\\
&=&
Z_0^B[0;T]  \exp  \left\{  \frac{1}{2} \intT \int_T d^2y \times \right.
\nonumber
\\
&&
\; \; \; \; \; \left.  J(x)
\Delta_T (x-y) J(y)\right\}.
\dla{freebosgenfun}
\ea
Here $N_{\beta}$ is an infinite  $T$-dependent constant arising in the 
 path integral description \dc{ber74}, the propagator is given in
(\dr{bosprop}) for small 
$\tilde{\mu}\beta$ and the free boson
partition function is $Z_0^B(T) = Z_0^B[0;T]$ as in
(\dr{partbos}).  Note that we have removed the $\tilde{\mu}$ labels on
propagator and partition function as there is no longer any possible
confusion with other mass scales.  Now define
$$
A_\pm=e^{\pm\left[i\lambda\phi\right]}
$$
(remember that we 
do not normal order the operators in path integral and so this
differs from the definition (\dr{Adef})).  Then
$$
\ll T_c \prodj A_+(x_j) A_- (y_j) \gg_0 \; =
\frac{ Z_0^B [\tilde J;T] }{ Z^B_0(T) } 
$$
with
$$
\tilde J(z) = - i \lambda \sum_{j=1}^n \left[ \delta^{(2)}(z-x_j)
-\delta^{(2)}(z-y_j)\right].
$$
As opposed to section \dr{sec:op}, here the UV
divergence of the propagator (\dr{bosprop}) is regulated by replacing
\be
 Q^2(0,0)\rightarrow  Q^2(\varepsilon_0,\varepsilon_1)=
 T^2 \varepsilon^2+\Od(\varepsilon^3)
\dle{bosonreg}
\ee 
where $\varepsilon_\alpha \rightarrow 0^+$ and 
$\varepsilon^2=\pi^2(\varepsilon_0^2+\varepsilon_1^2)$.  From
(\dr{freebosgenfun}),  (\dr{bosprop}) and  (\dr{bosonreg}) it follows
that
\ba
\lefteqn{ \ll T_c \prodj A_+(x_j) A_- (y_j)\gg_0= }
\nonumber
\\
&&
( T\varepsilon)^{n \lambda^2 /2\pi}\prodj  
\frac{  \prod_{j>k} \left[
Q^2(x_j-x_k)Q^2(y_j-y_k)\right]^{\lambda^2/4\pi} }
{\prod_{k=1}^n\left[Q^2(x_j-y_k)\right]^{\lambda^2/4\pi}}.
\dla{bosfreecorr}
\ea
The correlator is divergent due to the short-distance
(UV) divergent behaviour of the composite operator $\exp[i\phi(x)]$,
which needs to be renormalised.  We do this in the usual way through
the replacement of $\exp[ia\phi(x)]$, with $a\in
\IR$  arbitrary, by
\be
\left[\exp\left[i a \phi (x)\right]
\right]_{bare}=(\varepsilon \rho)^{a^2/4\pi}\left[
\exp\left[i a \phi(x)\right]\right]^R.
\label{renexp}
\ee
Here $\rho$ is an arbitrary renormalisation scale and the superscript
R will denote renormalised operators.  Note that (\dr{renexp}) is
analogous to (\dr{noexpphi}) with the identification
\be
\varepsilon = \frac{1}{\Lambda}
\dle{epsdef}
\ee
though in the operator formalism the renormalisation was carried out
through TNO.  Also observe that the $\rho$'s appear in the same way
though they have different origins---in the operator
approach
$\rho$ corresponded to an arbitrary mass at which normal ordering was
performed whereas here it is the arbitrary renormalisation scale.  From
equations (\dr{epsdef}) and (\dr{renexp}) observe that
(\dr{bosfreecorr}) reduces to (\dr{Acorrel}) as required.

In the full SG model one again works with the generating
functional which is expanded formally in powers of
$\alpha_0/\lambda^2$ \dc{GS}.  The
partition function $Z_{SG}(T,\mu = 0)$ is just obtained by
setting the external sources to zero.  We find \dc{GS} that with the
regularisation (\dr{bosonreg}) of the propagator, $\alpha_0$ is
renormalised just as in (\dr{alpren1}) for all the divergences to be
eliminated.  Once again the partition function is given by
(\dr{sgparfun}).

\subsubsection{MT model }

In calculating
the MT partition function in perturbation theory about $m_0 = 0$,
the correlator analogous to (\dr{Acorrel}) in the SG model is just
the TEV of insertions of the operators 
$\sigpm =\bar\psi (x)P_\pm \psi (x)$.  
Note, however, that the massless fermion theory is invariant under
chiral transformations 
$\psi\rightarrow\exp(i\alpha\gamma_5)\psi$. Under such transformation 
$\sigpm\rightarrow\exp (\pm 2i\alpha)\sigpm$ and therefore the
thermal average of a product of $\sigpm$ operators will vanish in 
the massless case unless the number of $\sigma_+$ and $\sigma_-$ is
the same. This is the  chiral selection rule, which only holds for
$m_0=0$.  Following \dc{zj}, the required correlator is obtained
by shifting
$\bar\psi\rightarrow \bar\psi\gamma^0$ so that,
naming  $\psi_a$ with $a=1,2$  the two components of the bispinor
$\psi$, the free massless theory decouples into two free
theories for the spinors $\psi_a$, and we have   
$\sigma_+\rightarrow \bar\psi_2\psi_1$ and 
$\sigma_-\rightarrow \bar\psi_1\psi_2$.  One obtains
\dc{GS,zj}
\ba
\lefteqn{
\ll  T_c  \prodj \sigma_+ (x_j) \sigma_- (y_j) \gg_0 =}
\nonumber
\\
&&
 (2\beta)^{-2n}\prodj 
\frac{\prod_{j>k}\left[
Q^2(x_j-x_k)Q^2(y_j-y_k)\right]}
{\prod_{k=1}^n\left[Q^2(x_j-y_k)\right]}.
\dla{ferfreem=0corr}
\ea
Notice that the above
correlator has exactly the same structure as the boson correlator
(\dr{bosfreecorr})---this is another peculiarity of 2D.  However,
unlike (\dr{bosfreecorr}), (\dr{ferfreem=0corr}) is finite since it
contains no product of fields at the same space-time point and there
are no mixing terms between 
$\psi_1$ and $\psi_2$  in the Lagrangian.

For the MT model one does have to worry about renormalisation (section
\dr{sec:op}).  The reason is that whilst the chiral symmetry is still 
unbroken so that
$\sigpm$ correlators still appear in the same number if 
$g\neq 0$ and $m_0=0$, now $(\bar\psi\gamma^\mu\psi)^2
\rightarrow 4\bar\psi_1\psi_1\bar\psi_2\psi_2$ when 
$\bar\psi\rightarrow\bar\psi\gamma^0$  and therefore there is
mixing between $\psi_1$ and $\psi_2$.  Thus products of fields at the
same point appear and the $\sigpm$ correlator becomes divergent: 
in the same way as the boson operator 
$\exp [ia\phi(x)]$, the $\sigpm$ composite operators need
renormalisation. Also as in the SG model, those are the only
infinities we have to worry about and they are absorbed in the
renormalised mass
$m$ as in (\dr{mren}) whilst
\be
\left[\sigpm\right]^R = (\varepsilon\rho)^{\kappa^2/\pi}
\left[\sigpm\right]_{bare}\label{rensigma}.
\dle{renbarm}
\ee 
Given these renormalisations, the MT model partition function is
obtained from the generating functional when the sources are set to
zero.  Its calculation requires some standard manipulations
\dc{fugasc82,na85,zj,Belg} (writing the quartic Thirring interaction
as a `gauge-like' interaction as well as calculating the 
 axial anomaly \dc{fuji84}) and one again obtains the
result (\dr{thparfun}).

\subsubsection{Zero-charge operators equivalences}\dle{sec:pi2}

So far we have shown the equivalence of the SG and MT partition
functions with $\mu =0$ using both operator and PI techniques. 
Further equalities between correlators of different operators in
each of the two models may also be shown to hold at $T>0$ and $\mu =
0$---we simply state the results here.  Further calculational
details may be found in \dc{GS}.

The equivalence of (\dr{bosfreecorr}) and (\dr{ferfreem=0corr}) in the
free massless bosonic and fermionic theories (after renormalisation)
may be extended to the SG and MT models where it becomes
\ba
\lefteqn{ \ll T_c \prodjN\sigma^R_+ (x_j)\sigma^R_-(y_j)\gg_{MT} \; = }
\nonumber
\\
&&
\left(\frac{\rho}{2}\right)^{2N}
\ll T_c \prodjN   A^R_+(x_j) A^R_- (y_j)\gg_{SG}.
\nonumber
\ea
We have also considered more complicated cases\footnote{For example
there could be unequal numbers of $\sigma_+$ and $\sigma_-$'s, since the 
 chiral
symmetry is broken in the MT model with $m_0 \neq 0$.}.  There the
calculation goes through in a similar way and leads to the expected
result in which  
\be
\sigma^R_{\pm} (x_j) \rightarrow
\left(\frac{\rho}{2}\right)  A^R_{\pm}(x_j).
\dle{repsig}
\ee

When dealing with correlators including insertions of the current
operator $j_{\mu}(x)$ in the MT model one is again faced with a
product of field operators at the same point, and hence with
additional divergences.  Using
point-splitting regularisation \dc{fried72} and taking particular
care to ensure that the Ward identities are satisfied, we have proved
that there are no extra divergences to renormalise and 
\ba
\lefteqn{ \ll T_c j^R_{\mu}(x) \sigma^R_+ (z_1) \sigma^R_-
(z_2) \gg_{MT} \; =}
\nonumber
\\
&&
\left(\frac{\rho}{2}\right)^2\frac{\lambda}{2\pi}\epsilon_{\mu\nu}
\ll T_c \partial_\nu^x\phi (x)
 A^R_+ (z_1) A^R_- (z_2)\gg_{SG}
\nonumber
\ea
where $j^R_{\mu}(x)$ is the regularised current \dc{GS}.  More
generally we find that the identities which hold between different
correlators seem to be simply obtainable through the replacements
(\dr{repsig}) and
\be
j^R_{\mu}(x) \rightarrow \frac{\lambda}{2\pi}\epsilon_{\mu\nu}
\partial_\nu^x\phi (x)
\ee
which are usually called operator bosonisation identities.   
 We stress, however, that firstly we do not believe these identities to be
strong---that is, we only expect them to hold between thermal
expectation values---and secondly it is not immediately obvious that
they hold inside all expectation values, though it does certainly seem
to be the case for the zero fermion charge ones we have considered.

\subsection{Bosonisation of the massive Thirring model at $T>0$, $\mu
\neq 0$}\dle{sec:chem}

Finally we have studied the Thirring model at nonzero chemical
potential $\mu$ for the conserved charge $Q_F = \int dx j^0(t,x)$
which is the net number of fermions minus anti-fermions;
$$  
{\cal{L}}_{Th} [\bar\psi,\psi;\mu] =  -\bar{\psi} (\pabar + m_0)\psi +
\frac{1}{2}g^2  j_\alpha (x) j^\alpha (x)+\mu j^0 (x)  ,
$$ 
and we have calculated the grand-canonical ensemble partition
function \dc{GS}.  Recall that now the averaged net fermion number
density
$\rho(\mu) = (\beta L)^{-1}\ll Q_F\gg$ is no longer necessarily zero,
and so a natural question to ask is  what is the bosonised version of
this theory. The answer was obtained in \cite{AlvGom98} for the
massless case where the free boson  partition function acquires an
extra 
$\mu$-dependent term.
In the massive case with partition function
$$
Z_{Th}(T,\mu)=N_\beta^F\pif \exp \left[-\intT 
{\cal{L}}_{Th} [\bar\psi,\psi;\mu]\right]
$$
we find in perturbation theory around the massless case, after 
using some of the  results in \cite{AlvGom98}  and performing some  
 functional integral manipulations, that 
$$
Z_{Th}(T,\mu)=Z_{SG\mu}(T,\mu)
$$
where
\ba
Z_{SG\mu}(T,\mu) &=& N_\beta \pib \exp\left[-\intT \left({\cal L}_{SG}
\right. \right.
\nonumber
\\
& &
\; \; \; \; \; \; \; \; \left. \left.
 -\frac{\mu\lambda}{2\pi}\frac{\partial}{\partial x^1}
 \phi(x)\right)\right].
\nonumber
\ea
(See \dc{GS} for a discussion of the boundary conditions.)  That is,
the bosonised action is the SG model plus an extra term which is
topological in that it only depends on the value of the field 
at the spatial boundary ($x=\pm\infty$).  From (\dr{Qdef}) this term is
interpreted as the result  of excitations with net kink (fermion)
charge being present in the thermal  bath and having associated 
chemical potential $\mu$ in the  grand-canonical ensemble.  Recall
that an analogous contribution was found in the chiral Lagrangian
for low-energy QCD \cite{AlvGom95}, where the r\^ole of kinks (Thirring
fermions) is played by the skyrmions (QCD baryons):  the chiral
Lagrangian at finite baryon density acquires, amongst other things, a
new factor $\mu Q_{SK}$ with
$Q_{SK}$ the skyrmion topological charge.

\section{Conclusions and Outlook}

After motivating the study of the SG and MT models at $T>0$
and $\mu \neq 0$, we have summarised some of the results obtained in
\dc{GS,GRS}.  Firstly, with zero chemical potential we were able to
show using both operator methods and path integral methods that the
partition functions were equivalent.  This also enabled a link to be
made between the arbitrary scale $\rho$ at which we carried out thermal
normal ordering in the operator approach and the arbitrary
renormalisation scale in the path integral---these are identical.
We then studied correlators of operators at different space time
points in each of the models.  Such results will be of crucial
importance for the application of this work to the estimation of the
number of topological (kink) defects formed in a phase transition, an
extension of this work which we motivated in the introduction
\dc{GRS}.  Also of relevance to this future work is the relationship
between the two models in the presence of a net number of
excitations;  we stated our results in sections \dr{sec:pi2} and 
\dr{sec:chem}.  

Finally, we are studying the link between the
sine-Gordon model and a 1D classical gas of positive and negative charges
with non-zero fugacity \dc{GRS}.  As a result of this (highly
studied system) and of the relationship between the sine-Gordon model,
Josephson junctions and the massive Thirring model, we would tentatively
suggest that such models should perhaps not be overlooked as ones in
which to develop or test calculational methods.
For example, one could try to investigate such difficult quantities 
as the pressure or the fermion number density, which could then give 
some insight into real physical problems.  
Similarly we hope to investigate the precise nature of the transition
at $\lambda^2 = 8 \pi$.

\section*{Acknowledgments}

D.A.S.\ thanks Peter Landshoff both for useful arguments and for
originally drawing her attention in this direction.  Thanks also to
Tim Evans for numerous helpful discussions, and to participants of {TFT
'98} for noting the conformal origin of the $Q$ variables.
M.Ogilvie has also pointed out some useful references. 
   Our
warmest thanks go to the organisers of {TFT '98} for an enjoyable and
stimulating conference and also a very useful pre-conference summer
school.  D.A.S.\ is supported by P.P.A.R.C.\ of the UK through a
research fellowship and is a member of Girton College, Cambridge. 
 A.G.N has received support through CICYT, Spain, project AEN96-1634 
 and through  a fellowship of  MEC, Spain,
  and wants to thank the Imperial College 
Theory Group for their hospitality during the completion of this work.  
This work was supported in part by the E.S.F.

\typeout{--- No new page for bibliography ---}

\end{document}